\documentstyle[12pt,epsf]{article}

\title{Zeno effect preventing  Rabi transitions \\
onto an unstable energy level}
\author{J\"urgen Audretsch\\
{\small Fakult\"at f\"ur Physik der Universit\"at Konstanz}\\
{\small Postfach 5560 M 674, D-78434 Konstanz, Germany}\\[2mm]
Michael B. Mensky\\
{\small P.N.Lebedev Physical Institute, 117924 Moscow, Russia}\\
{\small and}\\
{\small Fakult\"at f\"ur Physik der Universit\"at Konstanz}\\
{\small Postfach 5560 M 674, D-78434 Konstanz, Germany}\\[2mm]
Alexander D. Panov\\
{\small Russian Research Center ``Kurchatov Institute"}\\
{\small Kurchatov Sq., Moscow 123182, Russia}
}
\date{}
\newcommand{\eq}[1]{(\ref{#1})}

\newcommand{\Eq}{Eq.~\eq}

\newcommand{\Sect}[1]{Sect.~\ref{#1}}
\newcommand{\Sects}[2]{Sects.~\ref{#1},~\ref{#2}}
\newcommand{\Fig}[1]{Fig.~\ref{#1}}

\newcommand{\be}{\begin{equation}}
\newcommand{\ee}{\end{equation}}
\newcommand{\ba}{\begin{eqnarray}}
\newcommand{\ea}{\end{eqnarray}}
\newcommand{\ban}{\begin{eqnarray*}}
\newcommand{\ean}{\end{eqnarray*}}

\newcommand{\ra}{\rangle}
\newcommand{\la}{\langle}

\newcommand{\om}{\omega}
\newcommand{\Om}{\Omega}

\begin{document}
\maketitle

\begin{abstract}
We consider a driven 2-level system with one level showing
spontaneous decay to an otherwise uncoupled third level. Rabi
transitions to the unstable level are strongly damped. This simple
configuration can be used to demonstrate and to explore the quantum  
Zeno effect leading to a freezing of the system to the initial level. 
A comparison with repeated projection measurements is given. 
A treatment within a phenomenological theory of continuous 
measurements is sketched. The system visualizes the important 
role of null measurements (negative result measurements) and 
may serve as a good example for a truly continuous measurement. 
\end{abstract}

\section{Introduction}\label{intro}

The quantum Zeno effect
\cite{MisraSudar77-Zeno,ChiuSudarMisra77-Zeno,
Peres80-Zeno,Itano90-Zeno}
is one of the specific quantum effects which have gained much  
interest in particular in connection with the quantum theory of 
measurement. It goes back to successive or truly continuous 
measurements of the system and depends on the strength of 
the measurement. Generally speaking, it is the slowing down of  
induced quantum transitions as a result of repeated 
measurements of the system. In the limit of a continuous 
measurement, transitions of the system are 
inhibited. The system is frozen on a level.

\par
 
Characteristic for the Zeno effect is that it is caused by 
measurement. A prevention of a transition must not necessarily 
be a manifestation of the Zeno effect. It may also happen that a 
Zeno effect is accompanied by other influences. In the following 
we discuss a new demonstration of a pure Zeno effect.

\par

A well known particular example of the Zeno effect is the slowing 
down of Rabi transitions between two energy levels $|1\ra$ and 
$|2\ra$ of a $2$-level  system, which are induced by a resonant 
driving laser field, as the result of frequently repeated energy 
measurements. In the well-known experiment of Itano et al. 
\cite{Itano90-Zeno} the repeated energy measurement is 
realized for ions with the help of periodical short pulses of 
another laser field inducing optical transitions from level 
$|1\ra$ onto a third level $|3\ra$  with subsequent spontaneous 
decay back to the initial level $|1\ra$ ($V$-configuration). The 
photon emitted by spontaneous emission is monitored. If the 
ion is on level $|1\ra$, it will emit photons. This does not 
happen if it is on level $|2\ra$. In this sense the fluorescence 
photons contain  information about the state of the ion. The 
$|1\ra\rightarrow |3\ra$ transition together with the emission of 
the fluorescence photon represents a non-destructive 
projection measurement of level $|1\ra$. As a result the 
transition $|1\ra\rightarrow |2\ra$ is hindered. This freezing on 
a level is a pure demonstration of quantum Zeno effect caused 
by repeated almost instantaneous projection measurements 
(comp. \cite{Hegerfeld96}, where also a more complete list of 
the literature can be found).

\par

The aim of this letter is to show that the Zeno effect may be
demonstrated in a much simpler way for a driven $2$-level 
system with one level showing spontaneous decay to 
an otherwise uncoupled third level. At the same time this 
system may serve as a good example for a truly continuous 
quantum measurement. In addition it visualizes the important 
role of null measurements.

\par

We consider the very simple quantum system presented in   
\Fig{fig3-2level}a. It is  a $2$-level system with levels (energy
eigenstates) $|1\ra$ and $|2\ra$ which is subject to the influence 
of a resonant driving field $V$ generating Rabi oscillations 
between $|1\ra$ and $|2\ra$. Initially the system is  on level
$|1\ra$ and the driving field induces a transition to level
$|2\ra$. However level $|2\ra$ is assumed to be unstable. It 
decays rapidly by spontaneous decay with relaxation rate 
$\gamma$ to a third level $|3\ra$ with emission of a photon. 
Accordingly,  if the system reaches level $|2\ra$ it very quickly 
transits further to level $|3\ra$.  Therefore, if no photon is emitted, 
this is a sign that the system stays perpetually on level $|1\ra$.
Consequently we may consider this setup as a realization of a
continuous measurement of the energy level $|1\ra$ in a passive 
way leading to null-results (in contrast to the non-continuous and 
active scheme realized by Itano et. al. \cite{Itano90-Zeno}).

\par

What type of dynamical behavior is to be expected? Naively one could  
imagine that the system is transferred with the first Rabi pulse to level
$|2\ra$ from where it quickly decays to level $|3\ra$. The contrary  
turns out to be the case. The transition $|1\ra \rightarrow |2\ra $ is 
slowed down and the effectivity for this freezing on the initial level
$|1\ra$  actually increases  with the effectivity of the spontaneous  
decay of level $|2\ra$, i.e. with increasing relaxation rate $\gamma$.  
This may seem to be counterintuitive or paradoxical. But in fact it is
no surprise, because in a situation in which a continuous quantum
measurement is realized due to the coupling to the environment, 
the quantum Zeno effect is to be expected. We will show that
this behavior  of the system can  indeed very easily be understood  
as another experimentally accessible example of a pure manifestation 
of the Zeno effect. That this effect is in our case based only  i)~on 
null-measurements and ii)~on genuinely continuous  measurements,
makes it even more interesting from the conceptional point of view.

\par

We start in \Sect{System} with a quantum optical description of the 
microphysical dynamics underlying this Zeno effect. We compare it 
in \Sect{ComparProj} with the phenomenological approach based on 
repeated measurements. In \Sect{ComparRPI} we give a treatment 
within a phenomenological theory of continuous quantum 
measurements.  

\section{Quantum optical treatment of the system}
\label{System}

\begin{figure}[h]
\let\picnaturalsize=N
\def\picsize{1.5in}
\ifx\nopictures Y\else{\ifx\epsfloaded Y\else\input epsf \fi
\let\epsfloaded=Y
{\hspace*{\fill}
 \parbox{1.5in}{\ifx\picnaturalsize N\epsfxsize \picsize\fi  
\epsfbox{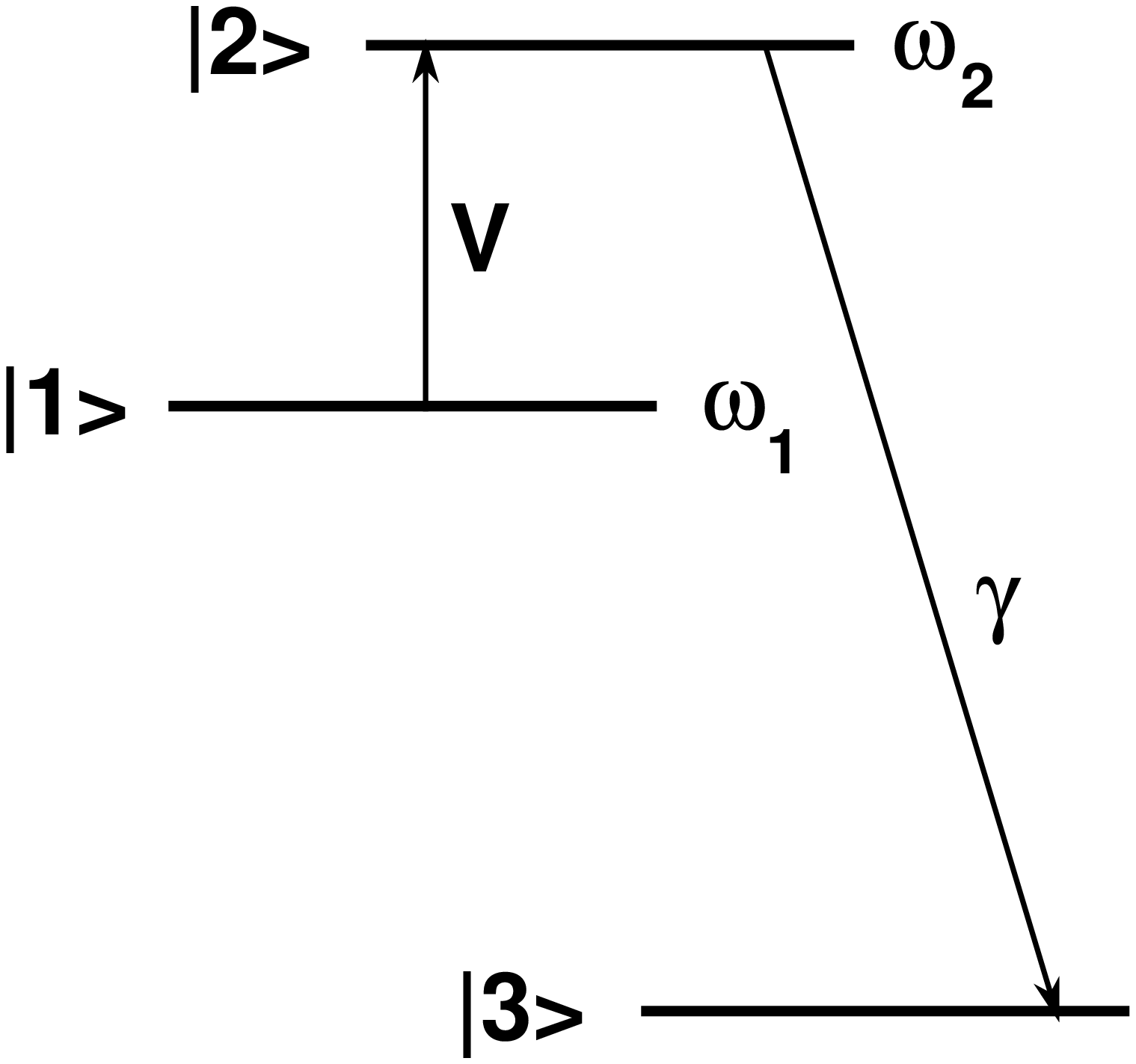}}\hfill
 \parbox{1.5in}{\ifx\picnaturalsize N\epsfxsize \picsize\fi  
\epsfbox{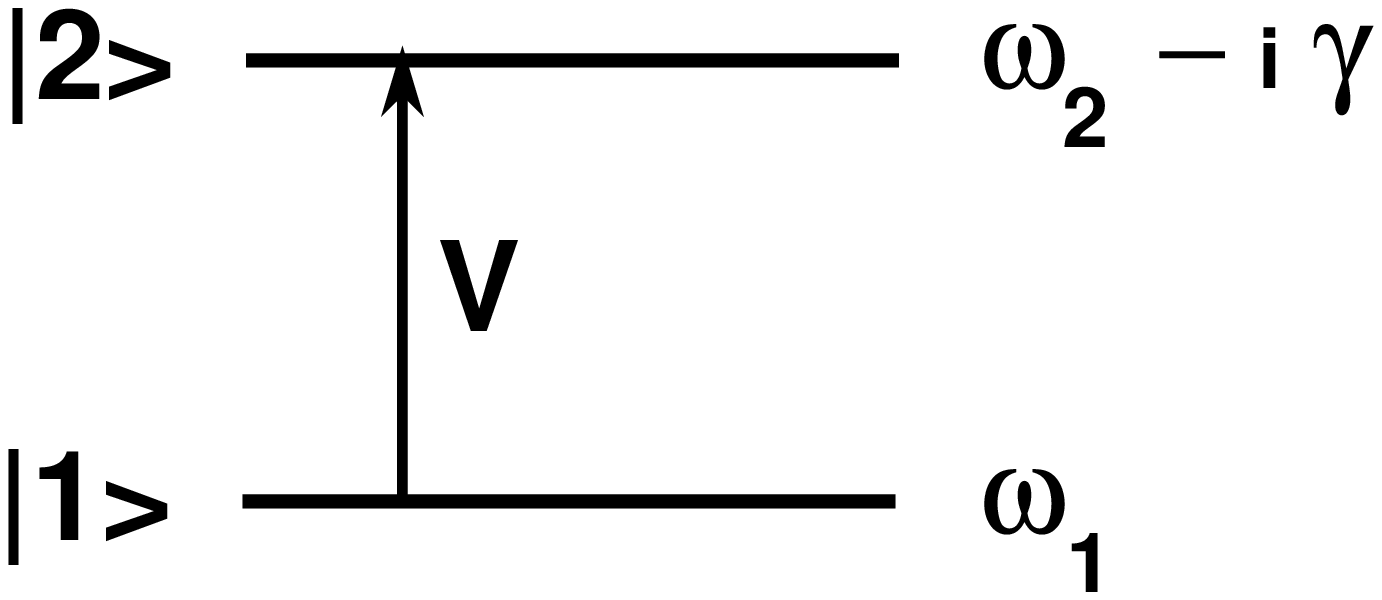}}\hspace*{\fill}}}\fi\\
\par\vspace{0.5cm}
\hspace*{\fill}$a$\hfill$b$\hspace*{\fill}\\
\caption{$3$-level system with initial state $|1\ra$ and Rabi
transition to state $|2\ra$ from where it only may decay to level
$|3\ra$  accompanied by the emission of a photon (Fig.~a). The  
equivalent $2$-level system has a complex energy of the 
unstable level $|2\ra$ (Fig.~b).}
\label{fig3-2level}
\end{figure}

We are  dealing with a 3-level system which is under the
influence of a driving field and in interaction with the
electromagnetic vacuum causing the spontaneous emission of a 
photon. An adequate  treatment of the complete underlying 
Schr\"odinger dynamics of the open system is  
rather involved. Let us therefore introduce
already on the level of the microphysical quantum optical treatment
some phenomenological elements. We study instead of the 3-level  
system an equivalent 2-level system consisting only out of the levels   
$|1\ra$ and $|2\ra$  with energies $\hbar \omega_1$ and 
$\hbar \omega_2$, and represent the instability of level $|2\ra$ by 
an imaginary term in the energy of this level (\Fig{fig3-2level}~b).

We write the Hamiltonian of the system as $H=H_{\gamma}+V$ 
with
\be
H_{\gamma}|1\ra =\hbar\om_1|1\ra, \quad 
H_{\gamma}|2\ra =\hbar(\om_2-i\gamma)|2\ra, \quad
\la 1|V|2\ra = \la 2|V|1\ra^{\ast}=\hbar\Om \, e^{i(\om_2-\om_1)  
t}\,.
\label{HamiltGamma}\ee
Let us  introduce amplitudes $a_1(t)$, $a_2(t)$ as follows: 
\be 
|\psi\ra=a_1(t)\,e^{-i\om_1 t}|1\ra+a_2(t)\,e^{-i\om_2 t}|2\ra\,. 
\label{amplitDefin}\ee
The dynamics  
\be 
|\dot\psi\ra=-\frac{i}{\hbar}(H_{\gamma}+V)|\psi\ra
\label{amplDenote}\ee
then result in the following differential equations for the  
amplitudes
\be 
\dot a_1=-i\Om a_2, \quad \dot a_2=-i\Om a_1-\gamma a_2\,.
\label{amplEq}\ee
{}
The solution is 
\ba 
a_1(t)&=&e^{-\frac{\gamma}{2}t}\left[ a_1(0)\cosh{\Om_\gamma t}
+\frac{1}{\Om_\gamma}\left( \frac{\gamma}{2}a_1(0)-i\Om  
a_2(0)\right)\sinh{\Om_\gamma t}\right], 
\nonumber\\
a_2(t)&=&e^{-\frac{\gamma}{2}t}\left[ a_2(0)\cosh{\Om_\gamma t}
-\frac{1}{\Om_\gamma}\left( i\Om  
a_1(0)+\frac{\gamma}{2}a_2(0)\right)\sinh{\Om_\gamma t}\right]\,, 
\label{amplSolInit}\ea
where we have introduced
\be 
\Om_\gamma=\sqrt{(\gamma^2/4)-\Om^2}\,.
\label{OmgDenote}\ee
Under the condition that the system is initially ($t=0$) on level
$|1\ra$, we find 
\be 
a_1(t)=e^{-\frac{\gamma}{2}t}\left( \cosh{\Om_\gamma t}
+\frac{\gamma}{2\Om_\gamma}\sinh{\Om_\gamma t}\right), 
\quad
a_2(t)=e^{-\frac{\gamma}{2}t}\left( 
-i\frac{\Om}{\Om_\gamma}\sinh{\Om_\gamma t}\right). 
\label{amplSolInitE1}\ee

\par

Depending on the relative strength of the two parameters
$\gamma$ and $\Omega$ one can discriminate different regimes of
behavior. If the influence of the driving field is strong as  
compared
to the strength of the spontaneous decay, i.e. for $\Omega\gg\gamma$,  
we
are in the Rabi regime, where the behavior of the system shows the
usual Rabi oscillations modified by some damping: 
\be 
a_1(t)=e^{-\frac{\gamma}{2}t}\cos{\Om t}, 
\quad
a_2(t)=-ie^{-\frac{\gamma}{2}t}\sin{\Om t}. 
\label{amplRabiReg}\ee
We see that the system, being initially on level 1, will be on level  
2 after the Rabi period $T_R=\pi/2\Om$.

\par

For our purpose it is the opposite regime with strong coupling
$\gamma\gg\Om$ to the environment which is of interest.
In this case the probability of permanent survival $P_{ps}(T)$ of the
system on level $|1\ra$ during sufficiently long time $T$ turns out  
to be: 
\begin{equation}
P_{ps}(T) = |a_1(T)|^2 = e^{-2\Om^2 T/\gamma}\,. 
\label{ProbSurvGamma}\end{equation}
Because of $\gamma \gg \Omega$ we find damping of the Rabi 
oscillations. In the limit $\Omega/\gamma \rightarrow 0$ we have 
$P_{\rm ps}(T) \rightarrow 1$ for any fixed $T$ and  the system is 
frozen on level $|1\ra$. This is a manifestation of  the Zeno effect. 

\par

We recall that the Zeno effect is the phenomenon that continuous 
measurement or many consecutive quantum measurements lead 
to damping and in a limit to freezing  of the evolution of the measured 
system. In our case we have a continuous measurement because  
of the coupling to the environment via spontaneous decay. And 
we have just shown that it results in freezing of the evolution of 
the system.

\par

It is a characteristic trait of our system that it is not closed, even 
when no photon is emitted. In fact this enduring failure to observe a 
photon in the interval $[0,T]$ represents  the continuum version of a 
series of nondetection measurements or negative result measurements. 
We called them null measurements. A null measurement influences the 
system in just the same way as the usual positive result measurement. 
It would be therefore misleading to call it  an ``interaction free 
measurement'' \cite{Dicke81,ElitzurVaidman93}. May be it should be 
better called an ``energy exchange free measurement'' 
\cite{KarlsonBjoerk98}. Our setup may serve as a simple example 
where the concept of null measurements leading to wave function 
collapses which result  in level freezing can successfully be applied. 
Other types of null measurements which are performed on the neutron 
spin \cite{PascazioNamiki93} or on the photon polarization 
\cite{KwiatWeinfurter95}  are discussed in the literature.

\par 

Finally we add that a full and detailed quantum optical treatment, which 
is not based on a complex energy as in \Eq{HamiltGamma} leads to the 
same result. It is therefore possible to obtain for our system the quantum 
Zeno effect within a fully microphysical approach. Note that it was not 
necessary to refer to a collapse of the wave function. We will now 
compare these results with the usual phenomenological approach to 
the Zeno effect, which is based on repeated measurements.   

\par 

\section{Comparison with repeated projection  
measurements}\label{ComparProj}

Let us first consider a situation for which the setup of Itano et. al. 
\cite{Itano90-Zeno} may be regarded as an experimental realization. 
We have a system with  two levels $|1\ra$ and $|2\ra$. We perform 
a sequence of strong measurements resulting in projections of the 
state vector (the continuous case will be discussed below). The two
corresponding  measurement results are $R_1$ and $R_2$. If $R_i$ 
is measured, the system is transferred to level $|i\ra$. Accordingly 
the measurement of $R_i$ is equivalent to the information that the 
system is immediately afterwards on level $|i\ra$. 

\par 

Let us assume that the  first measurement of a sequence is performed
at $t=0$ (thus fixing the initial state) and then repeated $N$ times with 
time interval $\tau$ between the consecutive measurements. $\tau$ 
is the relevant parameter. After the first repetition at time $t=\tau$ the 
system will still be found on the initial level with probability $q$ or on the 
other level with probability $p=1-q$. After the time $T=N\tau$ the
probability that the system has made $k$ times a transition to the
other level and has stayed $N-k$ times on the initial level is   
$P(k)=p^k q^{N-k}$. The probability that the system will be found 
finally at $t=T$ on  the initial level regardless of what have been the 
results of the measurements until then is   
\be
P^Z(T) = \sum_{k \;{\rm even}}\,{N\choose {k}} p^k q^{N-k}\,. 
\label{ProbSurv}\ee
 For any fixed $T$ the probability $P^Z(T)$   tends to $1$ for
$\tau\rightarrow 0$  (or $N\rightarrow\infty$). 

\par

For the experiment of the previous section the Rabi oscillation during 
the time $\tau$ fixes $p$ and $q$ to be 
\be
p = \sin^2\Omega\tau\,,\, q = \cos^2\Omega\tau\,.
\ee
To compare with the sequential approach sketched above, we have 
to take into account that our system is peculiar in the following sense: 
the first measurement result $R_1$ is ``no emission of a photon". 
This is a null result. We know that the system is on the initial level 
$|1\ra$. There is no alternative measurement result $R_2$  because 
``emission of a photon" is not connected with a projection onto  $|i\ra$, 
but indicates the end of the measurement series. Detection and 
emission of a photon are identified (see below). We know that all the 
measurements before have had the null result $R_1$. The related 
probability of permanent survival at level $|1\ra$ is obtained as the 
first term ($k=0$) of the sum \eq{ProbSurv}:
\be
P^Z_{ps}(T)= \cos^{2N}(\Omega\tau)\,.   
\label{ProbSurvOmegaC}\ee

\par

We turn now to the case of very rapid repetitions with $\tau\ll
\Omega^{-1}$. In this limit we may rewrite \Eq{ProbSurvOmegaC} 
without restriction of the time $T$ of permanent survival as 
\be
P^Z_{ps}(T)= \exp(-\Omega^2 T\tau)\,. 
\label{ProbSurvOmegaE}  \ee
Accordingly the result is that we can reach  complete agreement with 
$P_{ps}(T)$ of \Eq{ProbSurvGamma} by identifying
\be
\tau = \frac{2}{\gamma}\,.
\label{TimeSurv}
\ee
The repetition time $\tau$ must be chosen as the inverse of half of 
the relaxation rate. The quantum optical treatment of \Sect{System} 
uniquely fixes in our comparison the otherwise unspecified 
parameter $\tau$ . The condition $T\gg \gamma^{-1}$ of 
Sect.\ref{System} corresponds to the evident demand $T\gg \tau$. 
Perfect level freezing is obtained in the limit $\tau\propto \gamma^{-1} 
\rightarrow 0$.

\par

It is interesting to see that because of \Eq{TimeSurv}
the repeated collapse to the ground state happens  after time  
intervals $\tau$, which are completely fixed by the atomic lifetime. 
No reference to a photon detector is necessary. This demonstrates 
that the level freezing is independent of the detection or
nondetection of the emitted photon in a real photon-detector. Instead
it is the possibility of the irreversible decay into vacuum caused by
the interaction with the environment, which is responsible for the
level freezing .

\par

The fact that we are dealing with null measurements establishes 
another difference compared to the Itano et.al. experiment. There 
the probability has been investigated of finding the initial state at 
time $T$ regardless of the past history of the system in the interval 
$[0,T]$. This is to be described by $P^Z(T)$  given in \Eq{ProbSurv}.
But, as has been pointed out in \cite{NakazatoNamiki96}, this is not
the genuine quantum Zeno effect according to Misra and Sudarshan, 
which is based on the idea of permanent survival as it is expressed 
by $P^Z_{\rm ps}(T)$ of \Eq{ProbSurvOmegaC}. This needs 
continuous observations as they can be found in our proposal.  

\par

\section{Treatment within a phenomenological theory of continuous  
measurements}\label{ComparRPI}

The level freezing which we have observed above is a realization of 
a  genuinely continuous (strictly not sequential) measurement of 
energy, which has led to one uninterrupted null-result of duration 
$T$.  This  represents a particular measurement readout $[E]$, 
namely $E(t)=E_1={\rm const}$, with $0\le t \le T$. There are different
approaches to the theory of continuous quantum measurements 
(for a review see \cite{Mensky98}). 
We will refer to the presentation in \cite{AudMen97En} 
(for the realization scheme see \cite{AuMenEn98Realiz}). 
Note that the assumption of instantaneous wave function
collapse is not at all necessary in a phenomenological theory of  
continuous quantum measurement. Our phenomenological approach 
for example goes back to restricted path integrals.  We call it the 
method of the complex Hamiltonian (mcH). This fully elaborated 
scheme allows the treatment of all strengths of the influence of the 
measurement on the measured system from weak to strong.

\par

According to the mcH, a system with the Hamiltonian $H_0+V$, 
subject to the continuous measurement of energy resulting in the 
readout $[E]=\{E(t)\}$, is described by the Schr\"{o}dinger equation 
with the effective Hamiltonian containing an imaginary part: 
\be
H_{[E]}=H_0+V-i\kappa(H_0-E(t))^2, 
\label{ef-Ham}\ee
where the constant $\kappa$ characterizes the strength of the 
measurement. The inverse of this constant is a measure of the 
measurement fuzziness. 

\par

For comparison  with the  mcH we introduce a 2-level Hamiltonian  
$H_0$ with eigenvalues $E_i=\hbar\omega_i$ and rewrite our 
Hamiltonian $H_{\gamma}$ which describes the system without 
the driving field, in the form 
\be
H_\gamma=H_0-i\hbar\gamma |2\ra\la 2|\,.
\label{Hgamma}\ee
Then the complete Hamiltonian $H_{\gamma}+V $used in 
\Sect{System} for the description of our system is identical to 
\Eq{ef-Ham} provided that $E(t)\equiv E_1$ and the coefficient 
$\kappa$ is expressed by the level width $\gamma$ and the 
energy difference of the levels $\Delta E=\hbar(\om_2-\om_1)$ 
as follows: $\kappa=\gamma/\Delta E^2$. Notice that the 
measurement readout $E(t)\equiv E_1$ means in the mcH that 
the system, according to the measurement, is staying on the 
level $|1\ra$ all the time. Thus, there is complete agreement 
between the results of mcH and our dynamical consideration 
of \Sect{System}. 

\par

The so called level resolution time $T_{\rm lr} = 
1/(\kappa\Delta E)$ is a measure of the weakness of the
continuous measurement  and of the resulting fuzziness of the  
readout. It is in our case $T_{\rm lr} = \gamma^{-1}$. 
The condition $T_l\ll\Omega^{-1}$ characterizes what is called the
Zeno regime  (strong influence of the measurement).  This agrees 
with the condition used above in \Sects{System}{ComparProj}.   
The comparison with the phenomenological approach mcH has 
enabled us to introduce the concept of the weakness of our continuous 
null measurement and to fix it quantitatively. $H_\gamma$ refers to 
the particular readout $[E]=E_1$. According to the mcH the 
probability to obtain this $[E]$ agrees with $|a_1(T)|^2$ of
\Eq{ProbSurvGamma}.

\par

\section{Similar systems}\label{Analogues}

Some systems resembling the one considered above have been 
discussed in the literature.  

\par

In the papers \cite{Kraus81zeno,Chumakov96trapLevel} continuous
null measurements of transitions in a $2$-level ``atom"   
($|1\ra$,$|2\ra$) were considered. The atom occupies level 
$|1\rangle$ at the initial moment of time, its
evolution without observation is described by the equation
$|\dot\psi\rangle = -{\rm i}\beta\hat\sigma_1|\psi\rangle$, 
thus the atom oscillates between two levels with frequency $\beta$.
The atom is exposed to a continuous observation by an apparatus
indicating the transition from level $|1\rangle$ to level $|2\rangle$.
The apparatus is coupled only to the state $|2\rangle$ of the atom
by an interaction Hamiltonian of the form $|2\rangle\langle 2|\otimes
H$, where $H$ is a selfadjoint operator in the Hilbert space of the
apparatus. This coupling is assumed to switch on a process in the
apparatus resulting in the quick indication of the transition.
Neglecting the inner dynamical evolution of the apparatus, the  
authors showed that the evolution of the system is inhibited if the 
apparatus is sufficiently fast. Their conclusion is that the Zeno effect  
may take place and that it can be 
considered as a consequence of pure dynamical unitary evolution of
the combined system object-apparatus. The scheme of observation
considered in \cite{Chumakov96trapLevel} is however rather abstract
and no experimental realization of this scheme has been discussed. 

\par

The 3-level scheme identical to the one considered by us has  
been mentioned in the work of Plenio et al 
\cite{PlenioKnightThomps96decayZeno} in connection with the Zeno
effect but with regard to  a quite different aspect. Without any
detailed consideration, it was noted that the oscillatory behavior  
of probability $P_1$ is inhibited and replaced by the exponential one 
if level $|2\rangle$ spontaneously decays to another state during a
sufficiently short time. This feature of the system was related to a
finite width of the level $|2\rangle$, but the inhibition of oscillations
was not connected with the quantum Zeno effect. On the contrary, the
authors suggested to use the resulting behavior of level $|1\rangle$
as a model for a decaying system which shows a long enough
non-exponential period. This ``artificial" decaying system had then  
to be measured repeatedly to demonstrate the Zeno effect. 

\par

Rabi oscillations with radiative damping have been studied in  
different contexts. We give an example: The state
$|1\rangle$ is  considered to be the  ground state. A higher state
$|2\rangle$ is assumed to be unstable. The system decays back to 
the state $|1\rangle$ with emission of a photon. Investigation of such 
a system produces the well known result, that Rabi oscillations are  
damped by radiative decay. If the decay rate of level $|2\rangle$ is 
much greater than the Rabi frequency the probability to find the
system on the level $|1\rangle$ remains close to unity for all times. 
In this case (in contrast to our 3-level system) there are  
simultaneously two mechanisms of  ``freezing'' the system on the 
level $|1\rangle$: the fast transition from level $|2\rangle$ back to 
level $|1\rangle$ and in addition Zeno-like inhibition of Rabi 
oscillations connected with broadening of the level $|2\rangle$. 
Stimulated and spontaneous emission happen together. It is difficult 
to separate the two mechanisms. We cannot  interpret the resulting 
behavior as Zeno effect, because the two possible observational
results --- that a photon is or is not emitted --- are both related to the
same information, namely that the system is transferred to level  
$|1\ra$. This example  shows  that not any level ``freezing''  may
be taken as an indication for the presence of  a Zeno type  
measurement influence. In fact the analysis must be carried out  the 
other way round: If a situation can be interpreted as a measurement 
of Zeno type (including null measurements) than level freezing is to 
be expected.

\par

\section{Conclusions}
We have shown that the  freezing on  the initial level (inhibition of a 
transition) caused by the possibility of spontaneous decay of the 
other level can indeed be understood as a Zeno effect because it 
goes back to the measurement of the system. This is supported by a 
comparison with the phenomenological scheme based on repeated 
projection measurements and the phenomenologically described 
continuous energy measurement in the Zeno regime. After 
appropriate adjustment of parameters they all agree with the result 
of the quantum optical calculation. 

\par

The Zeno effect as described above is a particular quantum 
phenomenon related to quantum measurements which may occur 
in many different physical situations. To refer to it is of great 
predictive power: before any involved microphysical calculation 
has been performed, it must be expected that freezing on a 
level (or at least slowing down of a transition) will happen, if 
measurement influences the system. 

\par

Referring to our case, we know that quantum measurement may be 
explained as an
interaction with the environment. The spontaneous emission is the
mediator between the $2$-level system and the environment. It
suggests itself to make use of this by treating conversely the
influence of the environment as measurement of the system. This is
what we have done above.  As shown, the results are in accord with  
the microphysical calculation. The simplicity and the usefulness of 
the approach became evident.

\par

The advantage of the particular system presented above is, that it is  
easier to analyse than the standard example of Itano et al. 
\cite{Itano90-Zeno} and shows nevertheless to a certain extent a 
counterintuitive behavior. At the same time it is a very simple 
example of a null measurement, which is sometimes not quite 
correctly called interaction free measurement. Potentially the most 
interesting aspect of the system may be that it represents a truly 
continuous measurement, which can easily be handled and which we 
have discussed above only in its Zeno limit.

\vskip 0.5cm
\centerline{\bf ACKNOWLEDGEMENT}

This work was supported in part by the Deutsche 
Forschungsgemeinschaft and by the Russian Foundation for Basic  
Research, grant 98-01-00161. We thank Frank Burgbacher and 
Thomas Konrad for interesting discussions.


\begin{thebibliography}{1}

\bibitem{MisraSudar77-Zeno}
B.~Misra and E.~C.~G. Sudarshan.
\newblock {\em J.~Math. Phys.}, {\bf 18}:756, 1977.

\bibitem{ChiuSudarMisra77-Zeno}
C.~B. Chiu, E.~C.~G. Sudarshan, and B.~Misra.
\newblock {\em Phys. Rev.}, {\bf D~16}:520, 1977.

\bibitem{Peres80-Zeno}
A.~Peres.
\newblock {\em Amer. J. Phys.}, {\bf 48}:931, 1980.

\bibitem{Itano90-Zeno}
W.~M. Itano, D.~J. Heinzen, J.~J. Bollinger, and D.~J. Wineland.
\newblock {\em Phys. Rev.}, {\bf A~41}:2295, 1990.

\bibitem{Hegerfeld96}
A. Beige, G. C. Hegerfeldt and D. G. Sondermann.
\newblock {\em Quant. Semiclass. Opt.}, {\bf 8}:999, 1996. 

\bibitem{Dicke81}
R. H. Dicke.
\newblock {\em Am. J. Phys.}, {\bf 49}:925, 1981,
\newblock{\em Found. Phys.}, {\bf 16}:107, 1986.

\bibitem{ElitzurVaidman93}
A. Elitzur and L. Vaidman.
\newblock {\em Found. Phys.}, {\bf 23}:987, 1993.

\bibitem{KarlsonBjoerk98}
A. Karlson, G. Bj\"ork and E. Forsberg.
\newblock {\em Phys. Rev. Lett.}, {\bf 80}:1198, 1998.

\bibitem{PascazioNamiki93}
S. Pascazio, M. Namiki, G. Badurek and H. Rauch. 
\newblock {\em Phys. Lett. A}, {\bf 179}:155, 1993. 

\bibitem{KwiatWeinfurter95}
P. Kwiat, H. Weinfurter, T. Herzog, A. Zeilinger and M. A. Kasevich.
\newblock {\em Phys. Rev. Lett.}, {\bf 74}:4763, 1995.

\bibitem{NakazatoNamiki96}
H. Nakazato, M. Namiki, S. Pascazio and H. Rauch.
\newblock {\em Phys. Lett. A}, {\bf 217}: 203, 1996.

\bibitem{Mensky98}
M. B. Mensky.
\newblock {\em Physics Uspekki}, {\bf 41}:923, 1998. 

\bibitem{AudMen97En}
J.~Audretsch and M.~B. Mensky.
\newblock {\em Phys. Rev.}, {\bf A~56}:44, 1997.

\bibitem{AuMenEn98Realiz}
J.~Audretsch and M.~B. Mensky.
\newblock quant-ph/9808062. 

\bibitem{Kraus81zeno}
K.~Kraus.
\newblock {\em Foundation of Physics}, {\bf 11}:547, 1981.

\bibitem{Chumakov96trapLevel}
S.~M. Chumakov, K.-E. Hellwig, and A.~L. Rivera.
\newblock {\em Physics Letters A}, 197:73, 1995.

\bibitem{PlenioKnightThomps96decayZeno}
M.~B. Plenio, P.~L. Knight, and R.~C. Thompson.
\newblock {\em Optics Communications}, {\bf 123}:278, 1996.

\end{thebibliography}

\end{document}